\documentclass[12pt]{article}
\usepackage{graphics}
\begin{document}

\noindent
\begin{center}
{\bf \Large  Electromagnetic Dissociation and Space Radiation }
\end{center}

\begin{center}
{\large John W. Norbury} \\

{ \it Physics Dept., Worcester Polytechnic Institute, Worcester, Massachusetts, 01609, USA}
\end{center}
\begin{center}
{\large Khin Maung Maung} \\

{ \it Physics Dept., Hampton University, Hampton, Virginia, 23668, USA}
\end{center}

\noindent \hrulefill

\noindent {\bf Abstract}

Relativistic nucleus-nucleus reactions occur mainly through the Strong or Electromagnetic (EM) interactions. Transport codes often neglect the latter. This work
shows the importance of including  EM interactions for space radiation applications.

\noindent \hrulefill

\section{Introduction}

When astronauts travel into space they receive a significant dose of radiation because they are no longer protected by the Earth's atmosphere
and magnetic field. This radiation comes from three different sources, namely from geomagnetically trapped electrons and protons (Van Allen
radiation belts), from solar energetic particles (mainly protons) and from Galactic cosmic rays (protons and heavier nuclei) emitted by
stellar wind and flares and accelerated by supernova shock waves. The solar particles are particularly dangerous during times of solar flares
and coronal mass ejections.

A typical medical x-ray delivers a dose of about 0.1 mSv. The natural background radiation in the United States is about 4
mSv/year. The International Commission on Radiological Protection (ICRP) \cite{icrp}   has set a recommended
upper limit of an additional 1 mSv/year for the general public. The ICRP recommended upper limit for radiation workers \cite{icrp} is 20
mSv/year and for astronauts the recommended limit \cite{ncrp} is 500 mSv/year. Currently astronauts on
 the International Space Station (ISS)
 receive a dose of about 150 mSv/year, which 
is well below the maximum limit. However it has been estimated \cite{nasarp} that the dose
received on a Mars mission in a conventional spacecraft will be about 1000 mSv/year which is double the recommended limit for astronauts. The ISS
is in a low Earth orbit with an average height of only 400 km above Earth's surface and most of the orbit occurs at relatively low
latitude (low orbital inclination). Thus ISS still receives a significant amount of radiation protection from Earth's magnetic field. Most of
the dose received on ISS comes from traversal of the South Atlantic Anomaly and also during times when the orbit is at high inclination. The significant extra
dose of radiation that will be received on a Mars or Lunar  mission is due to the absence of a significant magnetic shield.

NASA has recently initiated a new program leading to human exploration of the Moon and Mars. 
Such long-term human missions will result in astronauts being exposed to the largest dose of space radiation ever experienced. 
 There is a
great need to provide accurate estimates of crew exposure to this radiation. 

A typical space radiation transport code works as follows. The
radiation environment outside the spacecraft needs to be simulated. This is done by using mathematical representations of the spectrum (i.e.
number of particles versus energy) for the various particles, such as electrons, protons, heavy ions etc. Of course the cosmic ray spectrum
is time dependent (especially as a function of solar cycle) and also varies depending on location in the solar system. These factors are also
included by some mathematical function. The spacecraft shield material (typically Aluminum) is also specified. The incident spectra are fed
into the Boltzmann transport equation, which takes the input spectrum, and converts it to an output spectrum, after radiation transport
through the spacecraft material. From the output spectrum one can then determine the radiation dose. Of course, for a Mars mission, one must
also include transport through the Martian atmosphere, just as one must include transport through Earth's atmosphere, when calculating
radiation doses experienced by high flying aircraft \cite{air}. The particles in the cosmic ray spectrum undergo atomic and nuclear
interactions with the particles in the shield material. These interactions are specified as atomic or nuclear interaction {\em cross
sections}. Thus the fundamental inputs to the Boltzmann equation are the incident cosmic ray spectrum and the particle interaction cross
sections. Once these are specified the Boltzmann equation can then be solved numerically either with deterministic  or Monte Carlo
 methods \cite{nasarp}. 

When one goes to the dentist, a lead (Pb) apron is donned during an x-ray. One might think that heavy materials such as Pb are good generic
radiation shield materials. Pb is certainly good for x-rays and electrons, but is very poor for protection against the more complex cosmic
rays spectrum which consists of protons and heavy nuclei. These nuclei interact with the Pb nuclei and because both projectile and target are
complex nuclei, there are many nuclear reactions which occur. In fact a {\em small} amount of Pb shield produces an increase, rather than a
decrease in the radiation that an astronaut receives. The same is true for the  Aluminum (Al) shields \cite{nasarp} from which spacecraft have
traditionally been made. One needs a {\em lot} of Al in order to provide adequate shielding. Lighter nuclei, such as hydrogen (H) or carbon
(C) are much more effective for shielding from cosmic radiation. From another point of view, aerospace engineers would always prefer to
construct aircraft and spacecraft out of light materials, in order to reduce heavy lift requirements. Thus it is indeed fortunate that light
materials provide the best protection against space radiation. This has been one of the major discoveries of the space radiation group at
NASA Langley Research Center \cite{nasarp}.

Most of the steady dose received on a Mars mission will come from radiation due to Galactic Cosmic Rays. There will also be transient periods
of high dose during times of intense solar activity. The Galactic Cosmic Rays are composed \cite{simpson} of about 98$\%$ nuclei and 2$\%$  
electrons and positrons. The nuclear component comprises about 87$\%$ protons, about 12$\%$ alpha particles and about 1$\%$ heavier nuclei.
Even though the heavy nuclei are not very abundant, they nevertheless contribute a large amount to the radiation dose because this depends on
$Z^2$ where $Z$ is the charge of the nucleus. The iron nucleus (Fe), being the most tightly bound of all atomic nuclei, is the most abundant
of the heavy nuclei in the cosmic ray spectrum and contributes a significant amount to the total dose. The number of nuclei heavier than Fe
drops off very rapidly, and are therefore of less importance.

The peak of the proton and heavier ion
flux in the cosmic ray spectrum occurs in the  energy region around 1 - 10 GeV. Fortunately this energy region is easily accessible to 
particle accelerators constructed over the last 50 years. This energy region is considered {\em intermediate energy}, with current
accelerators being able to also probe much higher and lower energy regions. Thus the cross sections needed as input to the Boltzmann equation
are able to be measured on Earth. 

The most important nuclear reactions that occur when a cosmic ray nucleus interacts with a nucleus in a spacecraft wall are called 
{\em nucleus-nucleus collisions}. For example the  incident projectile cosmic ray nucleus can be a proton or Fe nucleus and the target
spacecraft nuclei are typically aluminum  or light nuclei. If the projectile and target and nuclei heavier than hydrogen, the nucleus-nucleus
reaction is often referred to as a {\em heavy-ion reaction}. The four fundamental forces observed in nature are the Strong, Weak,
Electromagnetic and Gravitational forces. In a heavy ion reaction, the Weak and Gravitational forces can be neglected. The Strong force is
very short range, typically acting over distance of about a {\em fermi} (fm $ \equiv 10^{-15}{\rm m}$) and dropping to zero strength at larger
distances, whereas the weaker Electromagnetic (EM) force is significant over both short and long distances. At distances of the order of 1 fm
the Strong force completely dominates the EM force whereas the situation is reversed at distances significantly larger than 1 fm.
In a nucleus-nucleus collision, the nuclei might ``crash" into each other  or miss each other.  When they crash into each other (i.e. come closer than 1 fm), they undergo a Strong
interaction because of the small distance between them. When they miss each other (distance of closest approach larger than 1 fm), a Strong interaction will not occur, but they can
still interact via the longer range EM force. One can think of a photon traveling from one nucleus to the other and causing a nuclear excitation. 
There are many photons with low energy, dropping off to a few photons with high energy up to some maximum cutoff.
The most important photons are those with  frequencies near the resonant vibration frequency of the nucleus, which result in
 a nuclear excitation known as the Giant
 Dipole Resonance (GDR) where
entire nucleus undergoes large internal vibrations.  The GDR
 decays with the emission of nucleons.
Single nucleon emission is the most important, but multiple nucleon emission such, as two neutrons or alpha particles, 
are also significant.
Nucleus-nucleus reactions occurring via the Electromagnetic (EM) force are the subject of the present paper.
This reaction, with GDR excitation and subsequent nucleon emission, is called   {\em Electromagnetic Dissociation} and is often neglected in cosmic ray
transport codes. The aim of the present paper is to illustrate the importance of EMD for space radiation applications.

\section{Theory of Electromagnetic Dissociation}

In this section the theoretical description of EM dissociation 
 \cite{reanalysis, expl1, expl2, charge, single,
electric1, electric2, two, lifetimes, comment, higher, bertulani, llope} is reviewed.
One can feel the influence of a moving charged particle by the field surrounding it. In the quantum mechanical picture, the moving charged
particle is a source of virtual photons. The faster the particle moves, the more energetic will be the virtual photon field surrounding it. In
a nucleus-nucleus collision, when the nuclei miss each other, each nucleus can feel the influence of the other by the virtual photon field. To
be specific, when a projectile nucleus passes by a target nucleus, it feels the virtual photon field generated by the target, and these virtual
photons can excite the projectile (via excitation of the Giant Dipole Resonance) causing the projectile to subsequently emit nucleons. (The same
can also happen to the target from the photon field of the projectile.) Thus from the point of view of the projectile, it sees a beam of
photons emitted from the target and responds to these photons. Thus the fundamental reaction that the projectile undergoes is a {\em
photonuclear reaction} in which a photon is responsible for the excitation of the projectile.

 The traditional photonuclear process is a single high energy photon exciting a nucleus. Thus the whole field of photonuclear physics can be imported into our description of EM
dissociation. Actually in photonuclear physics it is typically an accelerated electron that provides the photons. In EM dissociation we have a
nucleus, instead of an electron, providing the photons.

How many photons does the target provide and what are their energies? This is described by the virtual photon spectrum $N(E)$, where $N$
is the number of photons with an energy $E$. A good description of this field is provided by the venerable Weiszacker-Williams (WW)
method of virtual quanta \cite{bertulani}, which gives the virtual photon spectrum 
\begin{eqnarray}
N(E) = \frac{2 Z_T^2 \alpha}{\pi E \beta^2} \Biggl\{ x K_0(x) K_1 (x) -\frac{1}{2} \beta^2 x^2 
\left[K_1^2(x) - K_0^2 (x)\right] \Biggr\} \label{ww}
\end{eqnarray}
where $N(E)$ is the number of virtual photons per unit energy $E$, $Z_T$ is the number of protons in the target nucleus, $\beta$
is the velocity of the target in units of $c$, and $\alpha$ is the Electromagnetic fine structure constant. The parameter $x$ is defined by
\begin{eqnarray}
x \equiv \frac{E b_{\rm min}}{\gamma \beta \hbar c}
\end{eqnarray}
where $\gamma = \frac{1}{\sqrt{1-\beta^2}}$ and $b_{\rm min}$ is the minimum impact parameter which is approximately equal to the sum of the
nuclear radii. 
$K_0(x)$ and
$K_1(x)$ are modified Bessel functions of the second kind.

The final EM dissociation cross section is written
\begin{eqnarray}
\sigma = \int  dE \;N(E) \sigma(E)   \label{sig}
\end{eqnarray}
where $N(E)$ describes the virtual photon spectrum provided by the target and $\sigma(E)$ describes the response (photonuclear
cross section) of the projectile to these target photons. (Again, the description can be reversed and the projectile provides photons which
excite the target.) 

An advantage of this Weiszacker-Williams formulation is that it is relatively easy to parameterize for use in radiation transport codes \cite{apj1, apj2}. Another big advantage
 is that {\em experimental} photonuclear cross sections can be used as input to equation (\ref{sig}).  When comparing this theory to EM
dissociation experiments, it is best to use such experimental photonuclear cross sections when comparing EM dissociation calculations to experiment. However
this is impractical when putting EM dissociation  calculations into transport codes. In that case one can use the following standard parameterizations of
photonuclear cross sections
\begin{eqnarray}
\sigma_{\rm abs}(E) = \frac{\sigma_m}{1 + \left[ (E^2 - E_{\rm GDR}^2)^2/E^2 \Gamma^2 \right]}
\end{eqnarray}
where $E_{\rm GDR}$ is the energy of the peak\cite{nim} of the GDR cross section, $\Gamma$ is the width of the GDR, and
\begin{eqnarray}
\sigma_m = \frac{\sigma_{\rm TRK}}{\pi \Gamma/2}
\end{eqnarray}
with the Thomas-Reiche-Kuhn cross section given by 
\begin{eqnarray}
\sigma_{\rm TRK} = \frac{60 N_PZ_P}{A_P} {\rm MeV mb}
\end{eqnarray}
where $N_P, Z_P, A_P$ are the neutron, proton and mass numbers of the projectile nucleus.  

The theoretical description provided in equations (\ref{ww}) and   (\ref{sig}) is able to provide a rough match between EM dissociation
experiments and theory \cite{reanalysis}. For much better agreement between theory and experiment, one must include a host of other
corrections such as electric quadrupole effects \cite{electric1, electric2}, Rutherford bending of the nuclear trajectories
\cite{expl2}, uncertainties in experimental photonuclear cross sections \cite{expl1}, correct separation of Strong and EM cross
sections in the original experiments \cite{expl2} and higher order effects \cite{expl1, lifetimes}.  When all
these effects are included one obtains good agreement with experiment \cite{expl2}.

\section {Nuclear Reactions for Space Radiation}

From a space radiation point of view some of the most important projectile cosmic ray nuclei are C, Si and Fe. Some of the most important
target (spacecraft shield) nuclei  are C and Al. The most important energies are in the 1 - 10 GeV range. Using the methods described above and
in the references \cite{reanalysis, expl1, expl2, charge, single,
 electric1, electric2, two, comment, higher}, I have calculated both Electromagnetic and Strong interaction cross sections for single
nucleon removal for a variety of projectiles and targets and energies. See Table 1.

Note that for light projectile-target combinations, such as a C projectile on a C target, the EM cross sections are very
much smaller than the Strong interaction cross sections at all energies. Therefore neglecting EM processes here is a good approximation. For
heavy projectile-target combinations, such as a Au projectile on an Fe target, the EM cross section is much bigger than the Strong interaction
process and therefore the EM interaction cannot be neglected.
The most important heavy ion reaction for space radiation is an Fe  projectile cosmic ray nucleus impinging on an Al target (spacecraft).
Depending on the energy the EM cross  section ranges from about 30$\%$  to 50$\%$ of the Strong interaction cross section! (At 50 GeV/A the
ratio is about 80$\%$.) Therefore {\em Electromagnetic dissociation should not be neglected in space radiation transport codes.}

\newpage
\noindent TABLE 1.  Calculations of Single Neutron removal from the Projectile. 

\noindent
\hrulefill
\begin{tabbing}
xxxxxxxxxxxxxx\=xxxxxxxxxxxx\=xxxxxxxxxxxxxxxxxx\=xxxxxxxxxxxxxxx\=xxxxxxxxxxxxxxxxxxx\=xxxxxxxxxxxxxxxx\=xxxxxxxx\kill
Projectile\> Target \>Tlab (GeV/A) \>$\sigma_{\rm Strong}$ (mb) \>$\sigma_{\rm EM}$ (mb)
\end{tabbing}
\begin{tabbing}
xxxxxxxxxxxxxx\=xxxxxxxxxxxx\=xxxxxxxxxxxxxxxxxx\=xxxxxxxxxxxxxxx\=xxxxxxxxxxxxxxxxxxx\=xxxxxxxxxxxxxxxx\=xxxxxxxx\kill
$^{12}$C      \>$^{12}$C    \>3     \>64      \>0.6         \\
              \>            \>5     \>64      \> 0.7        \\
              \>            \>10     \>64      \>0.9         \\
              \>            \>50     \>64      \>1.5         \\
              \>$^{27}$Al    \>3     \>77      \>2.4         \\
              \>             \>5     \>77      \>3         \\
              \>             \>10     \>77      \>4         \\
              \>             \>50     \>77      \>6         \\
              \>$^{56}$Fe    \>3     \>92        \>8.5         \\
              \>             \>5     \>92        \>11         \\
              \>             \>10     \>92        \>15         \\
              \>             \>50     \>92        \>25         \\
$^{28}$Si      \>$^{12}$C    \>3     \>73      \>1.1         \\
              \>            \>5     \>73      \> 1.4        \\
              \>            \>10     \>73        \> 1.8        \\
              \>            \>50     \>73        \> 3        \\
              \>$^{27}$Al    \>3     \>86        \>5         \\
              \>             \>5     \>86        \>6         \\
              \>             \>10     \>86        \>8         \\
              \>             \>50     \>86        \>14         \\
              \>$^{56}$Fe    \>3     \>100        \>17         \\
              \>             \>5     \>100        \>22         \\
              \>             \>10     \>100        \>30         \\
              \>             \>50     \>100       \> 52        \\
$^{56}$Fe      \>$^{12}$C    \>3     \>89      \>  7       \\
              \>            \>5     \>89        \> 8        \\
              \>            \>10     \>89        \> 11        \\
              \>            \>50     \>89        \> 17        \\
              \>$^{27}$Al    \>3     \>102        \> 27        \\
              \>             \>5     \> 102       \> 36        \\
              \>             \>10     \>102        \> 47        \\
              \>             \>50     \> 102       \> 77        \\
              \>$^{56}$Fe    \>3     \>116          \> 104        \\
              \>             \>5     \>116          \> 132        \\
              \>             \>10     \>116          \> 178        \\
              \>             \>50     \>116          \> 298        \\
\>\>\>\hspace*{2cm}  continued next page 
\end{tabbing}

\newpage
\noindent TABLE 1 continued

\noindent
\hrulefill
\begin{tabbing}
xxxxxxxxxxxxxx\=xxxxxxxxxxxx\=xxxxxxxxxxxxxxxxxx\=xxxxxxxxxxxxxxx\=xxxxxxxxxxxxxxxxxxx\=xxxxxxxxxxxxxxxx\=xxxxxxxx\kill
Projectile\> Target \>Tlab (GeV/A) \>$\sigma_{\rm Strong}$ (mb) \>$\sigma_{\rm EM}$ (mb)
\end{tabbing}
\begin{tabbing}
xxxxxxxxxxxxxx\=xxxxxxxxxxxx\=xxxxxxxxxxxxxxxxxx\=xxxxxxxxxxxxxxx\=xxxxxxxxxxxxxxxxxxx\=xxxxxxxxxxxxxxxx\=xxxxxxxx\kill
$^{197}$Au      \>$^{12}$C    \>3     \>128      \>  46       \\
              \>            \>5     \>  128      \>  56       \\
              \>            \>10     \> 128       \>73         \\
              \>            \>50     \> 128       \> 118        \\
              \>$^{27}$Al    \>3     \>141        \> 201        \\
              \>             \>5     \>141        \> 250        \\
              \>             \>10     \> 141       \> 330        \\
              \>             \>50     \>141        \> 541        \\
              \>$^{56}$Fe    \>3     \>156          \> 749        \\
              \>             \>5     \>156          \> 946        \\
              \>             \>10     \>156          \> 1263        \\
              \>             \>50     \>156          \> 2108        
\end{tabbing}

{\bf Acknowledgements}

 JWN was supported by NASA Research Grant NNL05AA05G. KMM acknowledges the support of  COSM, NSF Cooperative agreement  PHY-0114343.

\end{document}